\documentstyle[
prb,
aps]{revtex}
\begin{document}
\bibliographystyle{prsty}

\title{Scaling in the two-component surface growth}

\author{Miroslav Kotrla\cite{adr}}
\address{Institute of Physics, Academy of Sciences of the Czech Republic,\\
Na Slovance 2, 180~40 Praha 8, Czech Republic}

\author{Franti\v{s}ek Slanina}
\address{Institute of Physics, Academy of Sciences of the Czech Republic,\\
Na Slovance 2, 180~40 Praha 8, Czech Republic\\
and Center for Theoretical Study, Jilsk\'a 1, CZ-11000 Praha 1
, Czech Republic}

\author{Milan P\v{r}edota}
\address{Institute of Chemical Process Fundamentals,
Academy of Sciences of the Czech Republic,\\
165 02 Praha 6, Czech Republic
}

\date{\today }
\maketitle

\begin{abstract}
We studied scaling in kinetic roughening and phase ordering
during growth of binary systems using 1+1 dimensional single-step 
solid-on-solid model with two components interacting via Ising-like
interaction with the strength $K$.
We found that the model exhibits crossover 
from the intermediate regime,
with 
effective scaling exponents for kinetic roughening
significantly larger than for the ordinary single-step growth model,
to asymptotic regime with exponents of the Kardar-Parisi-Zhang class.
Crossover time and length are exponentially increasing
with $K$. For a given large $K$,
scaling  with enhanced exponents is valid over many decades.
The effective scaling exponents
are continuously increasing with $K$.
Surface ordering proceeds up to crossover.
Average size of surface domains increases during growth
with the exponent close to $1/2$,
 the spin-spin correlation function and the distribution
of domains obey scaling with the same exponent.
\end{abstract}

\vspace*{4mm}

\section{Introduction}

Growth by vapor deposition is an effective process for producing 
high quality materials. 
The microscopic mechanisms of growth were intensively studied in the past
mainly in the case of 
homoepitaxial growth \cite{levi97,vvedensky93,gilmer93}.
However, a common situation in nature as well as in modern technologies
is growth of binary or more component systems.
Due to nonequilibrium nature of growth the
properties of resulting film can be very different 
from properties of equilibrium bulk material, e.g., surface alloys 
which have no bulk 
analog can be formed or highly anisotropic structures can be prepared.
The problem of growth in a system with two or more components is
the problem of great practical importance but it is also
interesting from the pure statistical-mechanical point of view, because
growth process may belong to a new universality class
\cite{ausloos93,el-nashar96} and
such system might exhibit a nonequilibrium phase transition between 
low and high temperature phase.

There are two interfering phenomena in growth of binary systems: 
kinetic roughening 
and phase ordering.
During growth  the initially flat surface is becoming rough.
This is called kinetic roughening.
It has been found that this process often fulfills the invariance with respect
to scaling in both time and length.
Let us consider a surface in a $d$-dimensional space given
by a single-valued function $h({\bf r},t)$ of a $d^\prime$-dimensional
($d$$=$$d^\prime$$+$$1$) substrate coordinate ${\bf r}$.
The surface roughness is described by the surface width
$w(t,L)=\langle\sqrt{\overline{h^2}-\overline{h}^2}\rangle$,
where $t$ is the time, $L$ is a linear size
and the bar denotes a spatial average,
$\langle ... \rangle$ a statistical average.
It often obeys dynamical scaling law
$w(t,L)$$\propto$$L^{\alpha} f(t/L^{z})$, with the
scaling function $f(x)$ fulfills: $f(x)$ $=$ const., $x$$\gg$$1$
and $f(x)$ $\propto$ $x^{\beta}$, $x$$\ll$$1$ ($\beta$$=$$\alpha/z$).
Dynamical scaling allows to classify growth processes into
dynamical universality classes according to values of exponents
$\alpha$ and $z$ (or $\alpha$ and $\beta$)
\cite{barabasi95,krug97}.
This scaling has been
observed in a wide variety of growth models and 
many of them belong to the Kardar-Parisi-Zhang (KPZ)
universality class \cite{kardar86}. 
There has been considerable
effort in finding different possible universality classes. 

On the other side,
the process of ordering in ordinary (nongrowing) binary systems 
can lead to phase separation.
In the case of 
phase separation dynamical scaling exists as well, e. g. 
in the Ising model at low temperatures \cite{bray94}.
In phase ordering, the characteristic length $D$
is the average size of domains 
formed by particles of one type.
It increases with time as a power law, $D \propto t^{\psi}$.
The dynamics can be classified
according to values of the exponent 
$\psi$.
Phase ordering is
usually a bulk process, however, one can also
study ordering induced by growth only on the surface.
In this case the evolution of domain size
on the surface is of the interest.

On the microscopic level, growth is usually investigated using
discrete growth models.
Although several growth models 
for binary systems 
were introduced in various contexts,
e.g. for the study of phase separation during molecular beam epitaxy
 \cite{leonard97a},
or growth of binary alloys \cite{drossel97},
our understanding of growth of composite systems 
is still at the beginning. 
In particular, little is known so far about kinetic roughening 
in two-component
growth models.
This problem was probably first
considered by Ausloos {\it et al.} \cite{ausloos93}.
They introduced a generalization of the Eden model,
coined a magnetic Eden model,
which contains two types of particles with the probabilities of growth
given by the Ising-like interaction.
Ausloos {\it et al.} suggested that
the magnetic  Eden model does not belong to the KPZ universality class.
Recently El-Nashar {\it et al.} \cite{el-nashar96}
studied kinetic roughening in a ballistic-like two-component growth
model with the varying probability for deposition of given
type of particle. They observed that the exponent $\beta$
is changing with the varying probability and argued that kinetic roughening
no longer follows the KPZ scaling law.
Although the phase ordering was apparently present it was not studied
in these works.

In this paper we concentrate on the situation where both processes,
kinetic roughening as well as phase ordering, are important and 
affect each other.
We investigate scaling in both roughening of the surface and
phase ordering.
We use the one-dimensional
two-component single-step (TCSS) solid-on-solid growth 
model which we recently introduced \cite{kotrla97c}.
It is particularly convenient for the study of
the asymptotic scaling behavior.
Here we present results of extensive numerical simulations
which complement the preliminary results published elsewhere
\cite{kotrla97c,kotrla98a}.

The paper is organized as follows. In Sec. \ref{sec:model} 
 our model and the measured quantities are defined.
In Sec. \ref{sec:res} results of the Monte Carlo simulations
are described,
Sec. \ref{sec:dis} contains discussion.
Finally, a summary is given in Sec. \ref{sec:conc}.

\section{Model and measured quantities}
\label{sec:model}

\subsection{Modeling of two-component growth}
\label{sec:gen2c}
There is a large variety of the  single-component growth models
which can be potentially generalized to the multicomponent case.
Moreover, there are different possible ways of generalization.
One usually tries to use a model which is as simple as possible
and still contains relevant features.
Our aim is to find such model for study of scaling 
during two-component growth.

The  commonly used
approximation is application of a discrete model with 
the so-called solid-on-solid (SOS) 
condition.
It means that the surface is described by 
a single-valued function $h(i)$ taking discrete
values. The index $i$ is the horizontal coordinate 
which labels sites of the substrate. 
The rates for elementary growth processes depend usually only on 
values of $h$ in a neighborhood of the  initial and, in the case of
diffusion, possibly  also of the final 
position of a particle. 
The situation is more complex for two-component system
because the rates of elementary growth processes depend not only on 
the geometry but also on the local composition.
In practice it means that we need to store the composition in an
additional data field.
Let us  denote the type of a particle by a variable
$\sigma$ which takes values $+1$ or $-1$.
The geometry of the surface in time $t$ is described by the
function $h(i,t)$ and  the composition of the deposit is represented by
the function $\sigma(i,y)$, where  two variables $i$ and $y$ are
the horizontal and the vertical coordinates of a particle, respectively. The
variable $y$ is restricted only to values from 1 to $h(i,t)$.

Storing the composition of the whole deposit is possible
only for relatively small sizes of the substrate
and not too many monolayers (ML)
of deposited particles.
When we study the scaling phenomena, where the behavior for very long
times (i.e. many ML) is investigated, too much memory would be
required. 
However, when bulk processes can be neglected, it is sufficient to remember
only the composition within a certain finite depth under the surface,
because deeper layers cannot affect surface growth.  
The complication is that the depth which should be stored 
is in general not well defined and in principle it may be unlimited. For
example, to describe the rate for a process in which a particle is
moving from or to the bottom of a step we need to store the
composition in the depth equal to the maximum step size.
But it is known that in some models of MBE growth \cite{kotrla96a}
based on an unrestricted SOS model 
steps of an arbitrary size can be present.
Natural solution of this technical obstacle is to use the so-called
restricted SOS model 
in which  possible configurations are limited by an 
additional constraint $|h(i)-h(j)|\leq N$;
$i$ and $j$ being nearest neighbors and $N$ a given integer.

\subsection{Two-component single-step model}
Our model is based on the simplest restricted SOS model
which is the so-called single-step
solid-on-solid model.
The difference of heights between two
neighboring sites 
is restricted to $+1$ or $-1$ only.
The advantage of this choice is  that 
if we restrict ourselves  to nearest-neighbor interactions
between particles then we can
define rates for elementary moves of particles
using only the composition on the surface.
Hence, the rates at any time are given by the surface profile $h(i,t)$
and the composition {\it only} on the surface, which is described
by the field $\sigma(i,t)$  of the same dimensionality as $h(i,t)$.
We call such model {\it the two-component single-step} (TCSS) model.

Implemented growth rules depend in general on the physical situation under
study.
We consider rather simple case which, however, allows to evaluate
the effect of ordering on kinetic roughening.
As indicated above, we do not allow bulk processes which exchange particles.
This is well justified because rates for such processes are
usually several orders lower than for processes on the surface.
We also do not include surface diffusion.
This is a serious restriction from the point of application
to epitaxial growth. However, it is well known that
the study of scaling in models with the diffusion
is very computer power demanding already in the case of
one component growth \cite{kotrla96a} and that it is difficult
to obtain results with a good statistics.
We rather consider the condensation-evaporation dynamics.
Even more, we restrict ourselves here to the pure growth situation.
Evaporation can be included but we expect that it will
not change the scaling behavior provided the surface is moving,
i.e. deposition
occurs more frequently than evaporation. 

Hence, during the evolution, particles are only added.
Due to the single-step constraint a particle can be added only on the 
site at a local minimum of height, called growth site.
Once the position and the type of the particle are
selected, they are fixed forever.
The probability of adding a particle of type $%
\sigma $ to a growth site $i$
depends only on its local neighborhood 
and is controlled by a change of energy of the system
after deposition of a new particle.
The energy is given by the Ising-like interaction.
The probability is
proportional to $\exp \left\{-\Delta E(i,\sigma)/k_BT \right\}$,
where $k_B$ is 
the Boltzmann's constant, $T$ is temperature
and $\Delta E(i,\sigma)$ is the change of energy \cite{bkl}.

\begin{figure} 
\centering
\vspace*{100mm}
\includegraphics{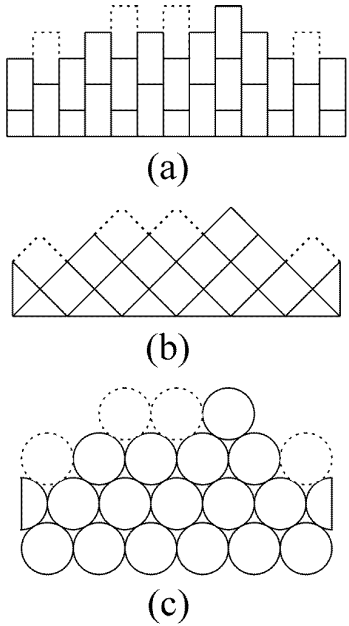}
\caption{Three realizations of the single-step geometry
in 1+1 dimensions. Dashed lines indicate positions where
a new particle can be deposited.
}
\label{fig:sssos}
\end{figure}

We describe explicitly our 
growth model for simplicity in 1+1 dimensions but it can
be straightforwardly generalized to any dimension.
Several realizations of the single-step geometry
are possible in 1+1 dimensions (Fig. \ref{fig:sssos})
\begin{figure}
\centering
\vspace*{40mm}
\includegraphics{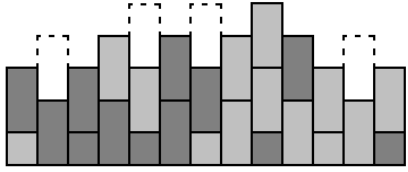}
\caption{Example of configuration with two types of particles
(dark and light gray) in the considered realization
of the single-step model. Dashed lines indicate positions where
a new particle can be deposited.
}
\label{fig:tcss}
\end{figure}
leading to three different variants of the TCSS model.
They differ in the number of
nearest neighbors of a new particle.
While in  the variant B there is an  ambiguity  in the type of newly
deposited particle
if two neighbors are of opposite types, this is not the case in variant A.
We expect that the effect of ordering on dynamics is stronger
for the variant A than for
the variant B.
The variant C (Fig. \ref{fig:sssos}c)
is technically slightly more complicated
to simulate due to the varying number of nearest neighbors of 
a deposited particle.
Therefore,  we consider the variant A
with three nearest neighbors which is represented
as stacking of rectangular blocks with the height equal to double of
the width (Fig. \ref{fig:sssos}a). 
Nevertheless, we expect
similar asymptotic scaling behavior for all three variants.

Then expression for the change of energy is
\begin{equation}
\frac{\Delta E(i,\sigma)}{k_BT}= -K\sigma [\sigma(i-1)+\sigma(i)+
\sigma(i+1) ] - H\sigma.
\end{equation}
Here, $K$ is a dimensionless coupling strength and $H$ is the bias
leading to preferential deposition of particles of  a selected type
($+1$ for positive, $-1$ for negative $H$). In analogy with magnetic
systems we will call $H$ external field.
The sum $\sigma(i-1)+\sigma(i)+
\sigma(i+1)$ contains
types of particles on the surface within nearest neighbors
of the growth site (which are three
in the chosen variant: left, bottom and right).

\subsection{Measured quantities and simulation procedure}
Evolution of the geometry is affected
by composition of the surface and vice versa.
We characterized the geometry of the surface by the
surface width $w(t,L)$ defined in introduction,
and by the
height-difference correlation function
$G\left( r,t\right) = \frac 1L\sum_{i=1}^L
\langle\left[ h\left( i+r,t\right)
-h\left( i,t\right) \right] ^2\rangle $,
which is expected to obey a scaling relation \cite{barabasi95,krug97}
$G(r,t) \propto r^{2\alpha} g(r/t^{1/z})$,
the scaling function $g(x)$ is constant
for $x$$\ll$$1$ and $g(x)$$\propto$$x^{-2 \alpha}$ for $x$$\gg$$1$.

In the case of phase ordering we concentrated on
evolution of the composition on the surface,
which is important for evolution of the geometrical profile 
of the surface.
We measured two quantities defined on the surface: 
i) the average of surface domain sizes, 
and ii) the correlation function $S(r,t)$ analogous to
spin-spin correlation function used in magnetic systems. 
We call the surface domain a compact part of the surface, which is
composed of particles of the same type. The size of the domain 
$d$ is measured along the surface.
Average size of the surface domains depends on time, 
on coupling, on external field and also on the initial 
composition of the substrate.
We denote the statistical average
of this quantity by $D=\langle d \rangle$.
The correlation function
$S(r,t)$ is defined as follows: $S(r,t)=\frac 1L\sum_{i=1}^L
\langle \sigma\left( i+r,t\right) \sigma\left( i,t\right) \rangle$.
The
only bulk property which we measured was concentration of 
particles of a given type,
$c_{+}$ resp.  $c_{-}$
(see subsection \ref{sec:magnetization}).

We performed simulations for various
coupling constant $K>0$
and mostly
for zero external field $H$ (except subsec. \ref{sec:magnetization}).
System sizes  varied from
$L=250$ to $L=80000$, and the number of deposited monolayers (ML) was up to
$3 \cdot 10^5$,
but for small systems even up to $4\cdot 10^6$ ML.
We measured  time $t$ of the simulation in the number of monolayers.
A statistical average was obtained by 
averaging over varying number of independent 
runs. It was from ten (for $L=80000$) up to several thousand 
(for $L=250$).

The growth starts on a flat surface configuration
\cite{flat}
as usual,
but in two-component models the evolution strongly
depends on initial composition of the substrate \cite{kotrla98a}.
Here we considered two possibilities 
i) a neutral substrate, i.e. substrate without any interaction with
deposited particles, in this case the system orders spontaneously
from beginning, and
ii) an alternating substrate, with the alternating types of particles.
The case of a homogeneous substrate composed of one type of particles
is reported elsewhere \cite{kotrla98a}.

\section{Results}
\label{sec:res}
\subsection{Evolution of morphology}
\begin{figure}
\centering
\vspace*{150mm}
\includegraphics{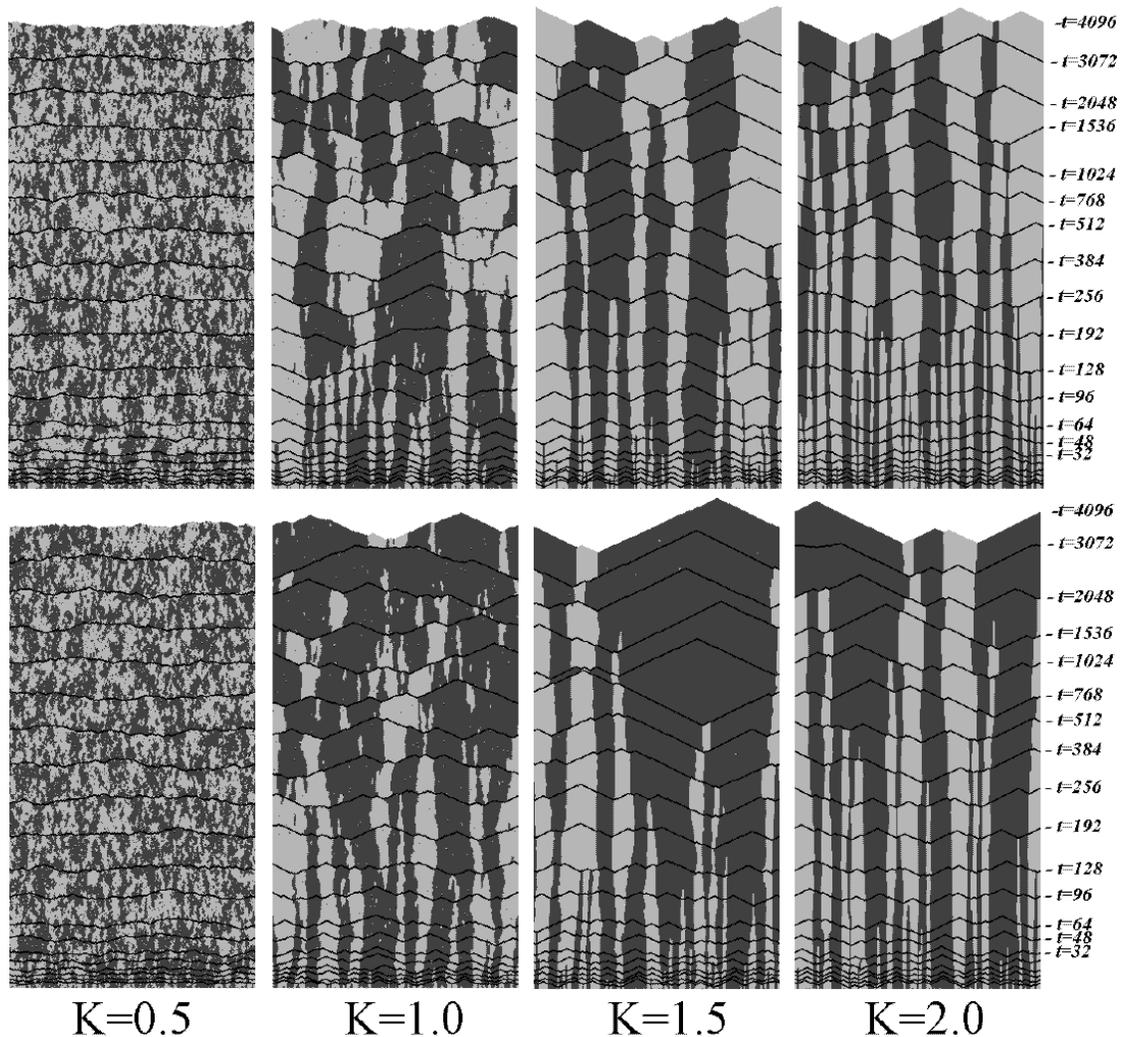}
\caption{Examples of evolution of surface profiles for
several values of coupling constant,
$K=0.5$, $1.0$, $1.5$, $2.0$,  and zero and nonzero external field $h$;
(upper panel - $H=0$, lower panel - $H=0.02$). 
Surface profiles at selected times increasing as powers
are indicated by black lines,
only part of the grown material close to the surface is shown at given time.
Dark and light gray correspond to different types of particles.
System size is $L=300$.}
\label{fig:conf}
\end{figure}

Fig. \ref{fig:conf} 
shows examples of time evolution
of surface morphologies and compositions  for 
selected couplings and external fields. Note that times
for which surface profiles are shown increase exponentially.
Visual inspection of many configurations leads to the following
observations.

In the case of zero external field (upper panel),
we can see that with increasing coupling the surface is becoming
more and more rough (faceted) and at the same time 
more and more clean columnar structures are formed;
the anisotropy induced by growth is more pronounced.
The average width of the columns increases with time.
We can also see that, for a given time,  average width of the column
is decreasing with coupling.
This is at the first sight in contradiction with the expectation
that ordering should be more pronounced for stronger coupling.
Notice also that for  larger coupling there is the correlation between
the domain walls and the local minima of the surface.
Both these effects can by explained from the dynamical rules of the
model. We shall discuss this point further in Sec. \ref{sec:dis}.

The lower panel demonstrates the 
effect of a small external field.
We can see that for small coupling  ($K=0.5$)
an external field does not cause
a significant change of the stoichiometry. The effect of the field is
canceled by fluctuations during growth.
However, for larger coupling
an external field leads to excess of one component. 
We evaluate this effect quantitatively in subsec. \ref{sec:magnetization}.

\subsection{Kinetic roughening}
The original single-step model belongs to the 
Kardar-Parisi-Zhang universality class \cite{kardar86}
with the exponents
$\alpha^{(KPZ)} =\frac{1}{2}$, $\beta^{(KPZ)} =\frac{1}{3}$, 
($z^{(KPZ)}=\frac{3}{2}$)
in 1+1 dimensions.
In this subsection we investigate the existence of  scaling
and the values of scaling exponents for kinetic roughening in the TCSS
model.

\subsubsection{Surface width}
\begin{figure}[hb]
\centering
\vspace*{100mm}
\includegraphics{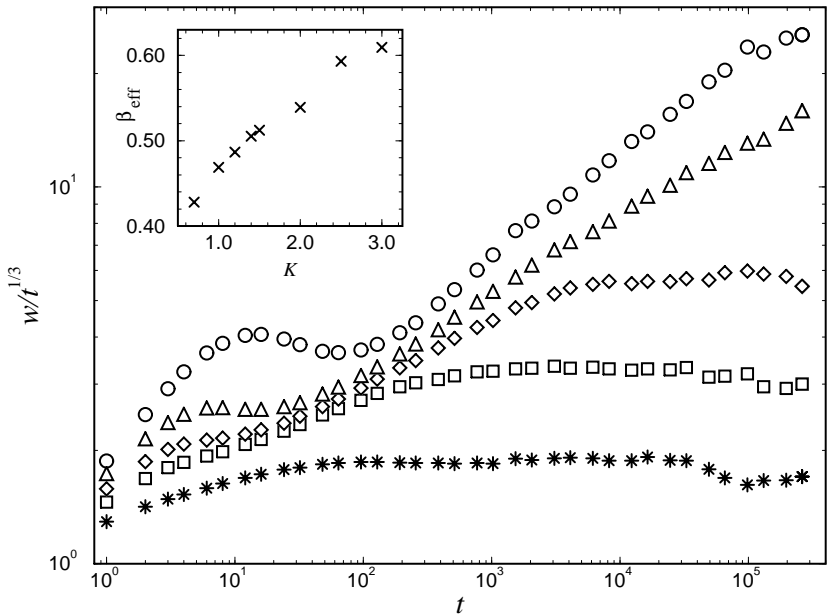}
\caption{Surface width $w$ divided by $t^{\frac{1}{3}}$ {\it vs.} time $t$ for
several values of coupling constant,
$K=$ 0.7 ($\ast$), 1.0 ($\Box$), 
1.3 ($\Diamond$), 2.0 ($\triangle$), 3.0 ($\circ$)
and zero external field,
$L=80000$.
Inset: Effective scaling exponent $\beta_{\rm eff}$ 
before crossover as function
of coupling.
}
\label{fig:rough}
\end{figure}

We start with
the time dependence of the surface width 
from which we can measure the effective exponent $\beta_{\rm eff}$.
In order to  avoid
the finite size effects caused by the saturation of the surface width
we used large system size $L=80000$.
We have found that the behavior changes with the
strength of coupling.
When the coupling is weak, evolution of the roughness
is almost the same as in the ordinary single-step model, e.g. for 
$K=0.3$ the increase of $w(t)$ can be well  fitted 
during all time of the simulation by one 
power law $w(t) \propto t^{\beta_{\rm eff}}$
with the exponent $\beta_{\rm eff}=0.33$, very close to the
KPZ value $\beta^{\rm KPZ} = \frac{1}{3}$ (cf. \cite{kotrla97c}).
For somewhat larger coupling
we have observed that the surface width exhibits 
the crossover in time.
At first the width increases with  an effective exponent
$\beta_{\rm eff} > 1/3$,
but after certain time $t_{\rm cross}$ it crosses over back
to $\beta^{(KPZ)}=\frac{1}{3}$.
This can be clearly seen
in Fig. \ref{fig:rough}
 (curves for $K=0.7$, $1.0$, and $1.3$) 
where we plotted the time dependence of the quantity
$w(t)/t^{1/3}$ for several 
couplings in order to compare evolution of the surface width with
the KPZ behavior.
Notice also that the absolute value of $w$ in a given time is increasing
with $K$.
The crossover from the regime with the enhanced $\beta_{\rm eff}$
to the regime with $\beta=\frac{1}{3}$ is definitely not 
a finite size effect. We checked that the time $t_{\rm cross}$
is the same for different system sizes.

\begin{figure}[hb]
\centering
\vspace*{100mm}
\includegraphics{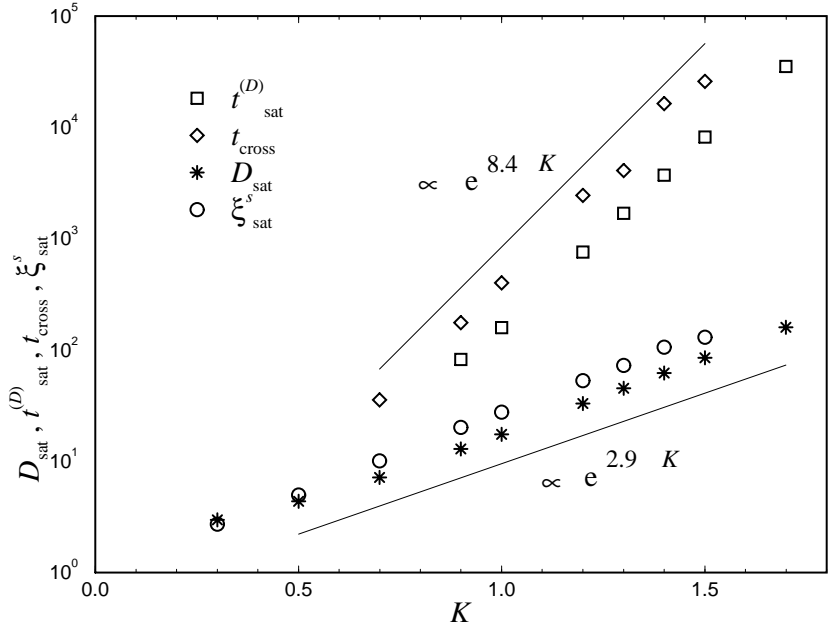}
\caption{Crossover time $t_{\rm cross}$ for the crossover in 
the behavior of the surface roughness ($\Diamond$),
saturation time $t_{\rm sat}$ for 
saturation of the average domain 
size ($\Box$),
saturated value of the average domain size $D_{\rm sat}$ ($\ast$), 
and saturated value of the correlation length $\xi^s_{\rm sat}$ ($\circ$)
{\it vs.} coupling $K$.
}
\label{fig:dep-on-K}
\end{figure}

From Fig. \ref{fig:rough} we can also see that $t_{\rm cross}$ 
is increasing with coupling.
We were not able to observe the crossover to the KPZ
behavior for coupling $K\ge 2$ in the time scale of our simulations
\cite{additional}.
In order to estimate the time needed 
we plotted the dependence of $t_{\rm cross}$ on $K$
(Fig. \ref{fig:dep-on-K}).
It can be fitted
as $t_{\rm cross}(K)\propto e^{8.4 K}$.
When we extrapolate this data to $K=2$ we get
 $t_{\rm cross}(K=2)>10^6$ which is longer
then the time of our simulation.

We found that for any $K>0.7$ there is a time 
interval extending over
several decades in which we can well fit our data as a power law
with the 
exponent $\beta_{\rm eff}>1/3$, before 
there is crossover to the KPZ 
behavior or our simulation stops.
The question remains, whether $\beta_{\rm eff}$ 
goes to a specific value for large
$K$. 
We did not find the indication that $\beta_{\rm eff}$
saturates to a certain value, at least for the investigated range
of coupling $K\in [0, 3]$.
The effective exponent $\beta_{\rm eff}$ is increasing function of $K$
(see inset in Fig.\ref{fig:rough}).
We attribute the rather large value of $\beta_{\rm eff}$ 
for strong coupling to pinning of the surface at domains boundaries.

\subsubsection{Height-difference correlation function}
The second exponent for kinetic roughening is the roughness exponent
$\alpha$.
It can be calculated from the dependence of the saturated
surface width 
on the system size. We used here an alternative and often more
accurate way. We calculate $\alpha$ from the spatial dependence
of the height-difference correlation function
$G\left( r,t\right)\propto r^{2\alpha}$ in the long time limit.

\begin{figure}[hb]
\centering
\vspace*{90mm}
\includegraphics{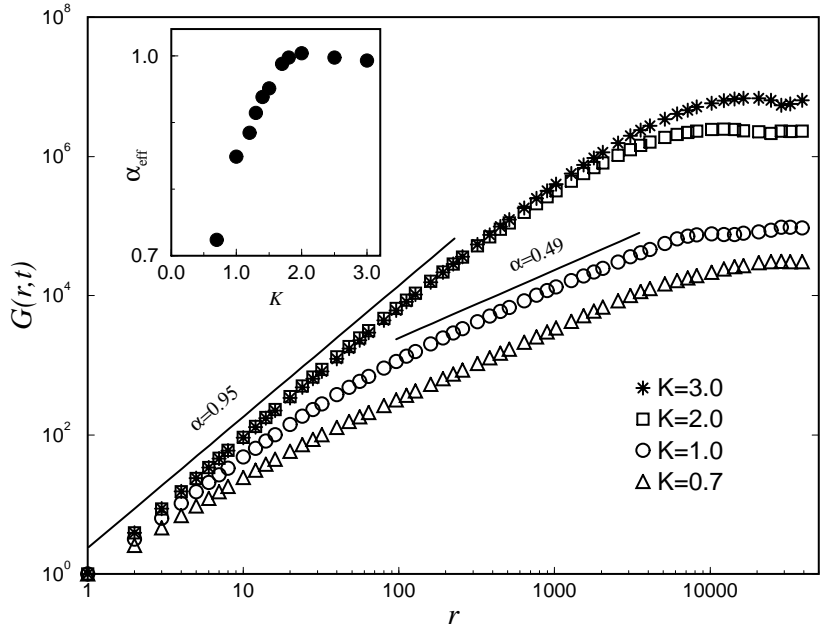}
\caption{
The height-difference correlation function
$G(r,t)$ for couplings $K=0.7$, $1.0$, $2.0$, and $3.0$ 
after 262144
ML were deposited, $L=80000$.
Inset: Effective exponent $\alpha_{\rm eff}$ before crossover 
as function of coupling.
}
\label{fig:hcor}
\end{figure}

The obtained exponents $\alpha$ also depend on coupling. 
Exponent $\alpha$ has a value close to
$\alpha^{(KPZ)}=\frac{1}{2}$ for weak coupling.
For larger coupling,
we have found that when the system is sufficiently large there is
a crossover behavior in the spatial dependence of 
the height-difference correlation function (see Fig. 
\ref{fig:hcor}).
On the small length scale it increases faster 
than $r$, and for sufficiently large $K$ we can fit
it as a power law $r^{2\alpha_{\rm eff}}$
 with an effective exponent  $\alpha_{\rm eff}> \frac{1}{2}$.
However, if the distance is larger than certain length $l_{\rm
cross}$, the form of the function crosses over to power law with
the exponent which is close to 
$\frac{1}{2}$,  i.e.  to the KPZ behavior
(cf. data for $K=1$ in Fig. \ref{fig:hcor}).

If the coupling is strong $(K\ge 2)$
we  do not see the crossover
but only the larger exponent $\alpha_{\rm eff}\approx1$.
We expect that it is because
even the system size $L=80000$ and time 262144
ML are not large
enough to get into the crossover regime.
We have found that $\alpha_{\rm eff}$ 
is increasing smoothly 
with the strength of coupling
from $\alpha^{(KPZ)}=\frac{1}{2}$ to a
value slightly less than one (inset in Fig. \ref{fig:hcor}).
The value $\alpha=1$ is the natural limit 
imposed by the single-step constraint.

\begin{figure}[hb]
\centering
\vspace*{90mm}
\includegraphics{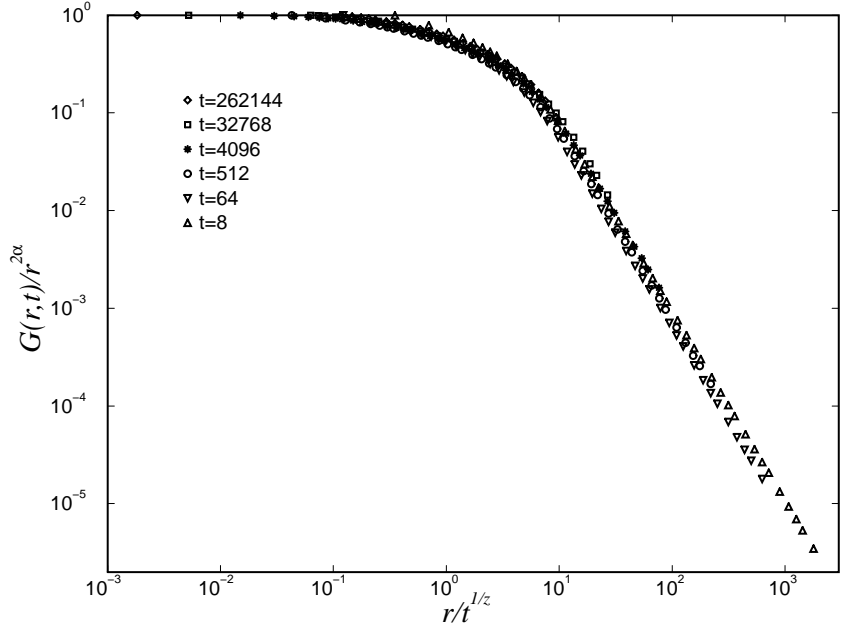}
\caption{
Data collapse of the correlation function $G(r,t)$
for coupling $K=2$ obtained for exponents $\alpha =0.97$ and $z=1.97$. 
}
\label{fig:collapse}
\end{figure}

\subsubsection{Scaling}
Having both exponents $\alpha$ and $\beta$, we can try to verify scaling.
For weak coupling, the exponents are close to the KPZ exponents
and scaling with these exponents is satisfied.
For larger coupling, when the crossover is observed,
we cannot get scaling for all times and lengths,
nevertheless for long times and large lengths
the KPZ scaling is valid.
When coupling is sufficiently strong
then  the behavior characterized by
enhanced exponents extents over many decades of time and length 
and looks
practically as asymptotic.
Then we can ask ourselves if there is scaling with new exponents
which is satisfied on this scale.
In order to show that there is scaling, 
we should get data collapse.
We have found that indeed we get the data collapse
over many decades in the strong coupling regime ($K\ge 2$).
As an example we show in Fig. \ref{fig:collapse}
the data collapse  of the height-difference correlation function
$G(r,t)$ obtained for 
$K=2$ with exponents $\alpha=0.97$ and $z=1.98$.
For different $K$ we need different exponents, 
e.g., for $K=3$ we get the best data collapse for  $\alpha=0.99$ and $z=1.7$.
The question remains what is true asymptotic scaling in the
strong coupling regime. 
We expect that for any $K$ there is crossover to the KPZ scaling
(although $t_{\rm cross}$ may be astronomically large),
and the asymptotic behavior will belong to the KPZ class.

\subsection{Phase ordering}

\subsubsection{Time evolution of surface domains}
\begin{figure}[hb]
\centering
\vspace*{90mm}
\includegraphics{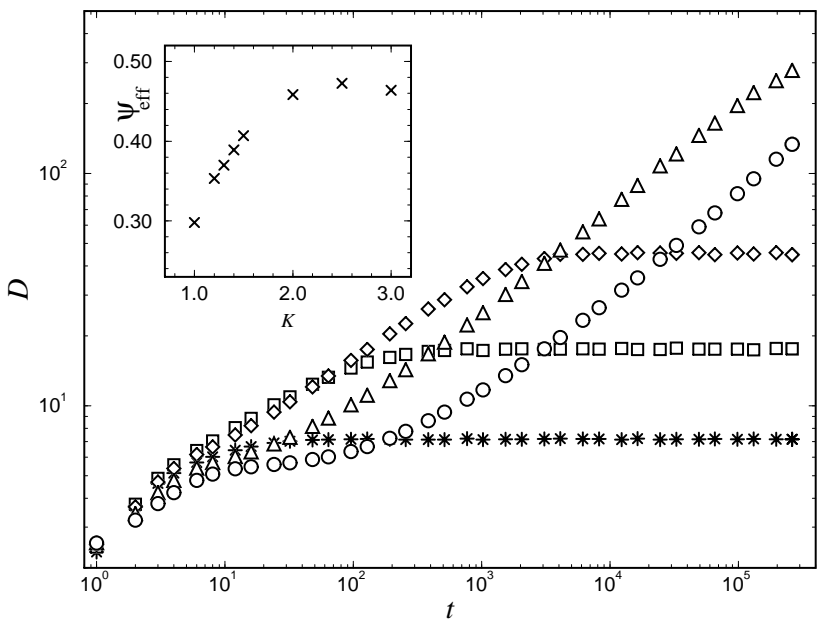}
\caption{
Time evolution of average surface domain size for
coupling constants $K=$ 0.7 ($\ast$), 1.0 ($\Box$), 
1.3 ($\Diamond$), 2.0 ($\triangle$),3.0 ($\circ$)
and zero external field, $L=80000$.
Inset:
Effective  exponent $\psi_{\rm eff}$ as function of coupling $K$.
}
\label{fig:d-ev}
\end{figure}

Time evolution of the average surface domain size $D$ for several
couplings is shown
in Fig. \ref{fig:d-ev}.
We can see that the behavior again depends on $K$.
For small coupling, $D$ at first increases, 
however after some time $t^{(D)}_{\rm sat}(K)$ surface domain size saturates
to a value $D_{\rm sat}(K)$.
We have checked that the saturation
is not a finite size effect (see \cite{kotrla97c} - inset in Fig. 3).
This is  an intrinsic property of the model.
Both  the time $t^{(D)}_{\rm sat}(K)$  and the saturated value
$D_{\rm sat}(K)$ rapidly increase  with $K$ (Fig. \ref{fig:dep-on-K}).
We have found that the dependence on $K$ can be well fitted by
exponentials:
$D_{\rm sat}(K) \sim {\rm e}^{(2.91\pm0.06)K}$ and
$t^{(D)}_{\rm sat}(K) \sim {\rm e}^{(7.67\pm0.06)K}$.
We were not able to observe saturation for $K=2$ 
because of prohibitively long simulation time needed.
Hence,
for $K\ge 2$ the domain size is increasing
during all time of our simulation.
But we believe that the evolution of the surface domains
is analogous to evolution of domains in one-dimensional Ising model
and that in long time it will saturate for any $K$
\cite{kineticIM}.

Evolution of domains is strongly affected by initial conditions
and a certain transition time is needed before a value independent on the
initial state is reached. This phenomenon is similar to what we have
observed in the case of surface width.
The transition time is increasing with coupling 
and can be quite long, e.g. for $K=3$ it is several hundred of ML.

We have found that for $K\ge 1$ there is a time interval in
which the increase of average domain size
can be fitted by a power law, $D(t,K)\propto t^{\psi_{\rm eff}(K)}$ 
with an exponent $\psi_{\rm eff}$ depending on $K$.
For large $K$ the exponent $\psi_{\rm eff}$ seems to saturate to
a value slightly smaller than $\frac{1}{2}$
(see inset in Fig. \ref{fig:d-ev}).
This is the same exponent as for the Ising
model with nonconserved order parameter \cite{bray94}.

\subsubsection{Distribution of surface domain sizes}
The average surface domain size contains the information about the 
formation of the domains during the growth. 
However, it is not
clear only from that quantity, whether the domains form a kind of
periodic structure with a typical domain length or the domain sizes
are  rather random. In order to obtain more information about the
domains, we 
measured the probability distribution $P(d,t)$ of domain sizes $d$ 
as a
function of time. We performed the simulations 
for three values of coupling $K=1$, $K=2$ and $K=3$,
the system size was $L=1000$, and time up to 32768
ML.
In order to get good statistics we had to make
the average over 
ten thousand independent runs. 
We observed that for initial times there is a rather sharp 
asymmetric peak
with the position shifting to higher values of $d$ with increasing time.
During time evolution the amplitude of the peak  decreases
and the  peak becomes more and more broad and eventually disappears.
The time scale for this behavior depends on $K$.

We try to rescale our data extending  over many orders of magnitude
and to look if there is  scaling.
We applied
a scaling formula of the form
\begin{equation}
P(d,t)= F(d/D(t))/D(t),
\label{eq:pofdscaling}
\end{equation}
with a scaling function $F$.
The average domain size $D(t)$
is in fact equal to the average computed from the
distribution $P(d,t)$.

\begin{figure}[hb]
\centering
\vspace*{90mm}
\includegraphics{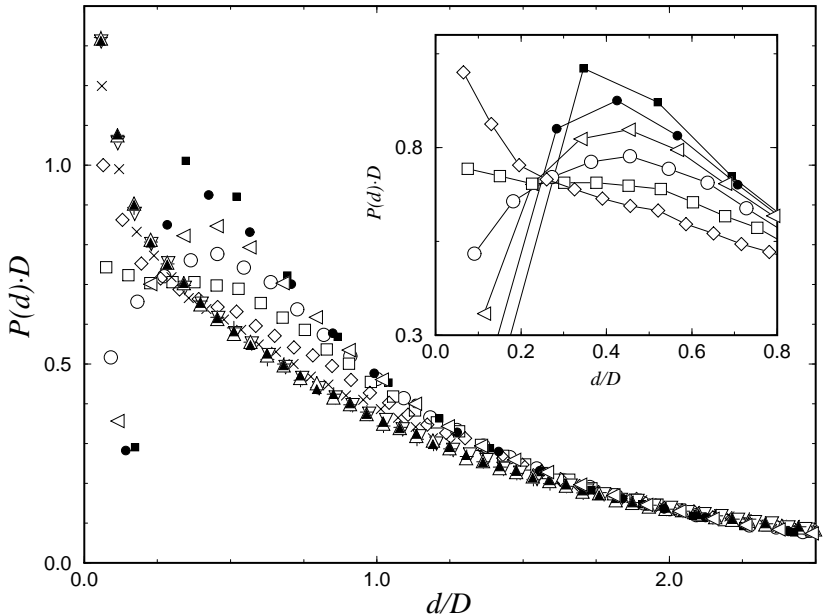}
\caption{
Distribution of domain sizes for $K=1$ and times
4
(full square),
8
(full circle),
16
(triangle left),
32
(circle),
64
(square),
128
(diamond),
256
(times),
512
(triangle down),
1024
(plus),
6144
(triangle up),
32768
(full triangle).
Inset: Detail of the distribution. The symbols are the same as
in the main graph.
}
\label{fig:dsd2}
\end{figure}

In the Fig. \ref{fig:dsd2}
 the rescaled distribution of
domain sizes for $K=1$ is shown as a function of the variable
$x=d/D$, for times from 4 to 32768.
We can see that the peak at around $x=0.4$ exists only for short times
and for time 
about $t= 64 $ it changes to a small plateau which further vanishes
for longer times. The distribution of domain sizes for long times
converges to a function
which we found to be well fitted by an exponential. 
Moreover, we found that the peak vanishes around the time 
$t^{(D)}_{\rm sat}$ when the
average domain size saturates.

\begin{figure}[hb]
\centering
\vspace*{90mm}
\includegraphics{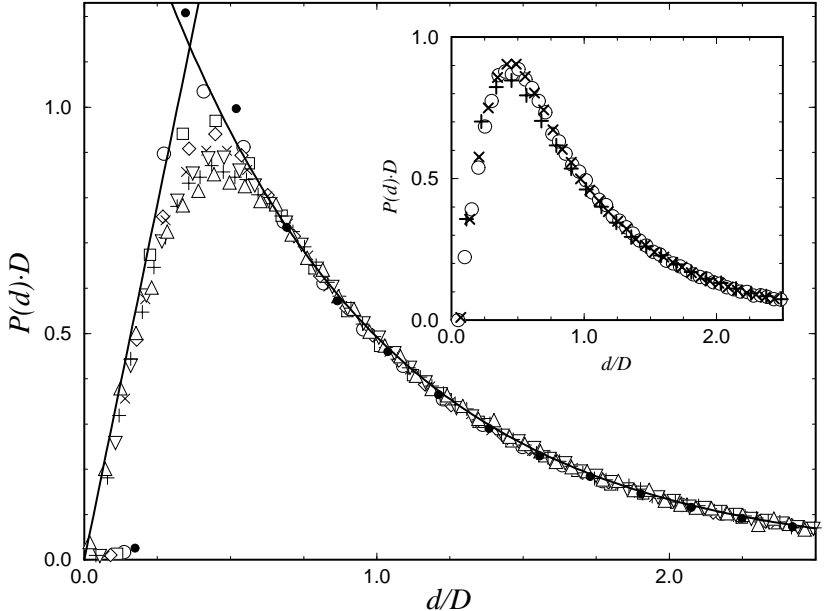}
\caption{
Distribution of domain sizes for $K=2$ and times
8
($\bullet$),
32
($\circ$),
64
($\Box$),
128
($\Diamond$),
256
($\times$),
512
($\bigtriangledown$),
1024
(+),
6144
($\triangle$).
The full lines are exact results for the kinetic Ising model,
ref. [21], in the limits of small and large $d/D$.
Inset: Distribution of domain sizes for $K=1$ in time
16 (+), for $K=2$ in time 
256 ($\times$) and for $K=3$ in
time 
4096 ($\circ$).
}
\label{fig:dsd1}
\end{figure}
The results for $K=2$ are shown in Fig.
\ref{fig:dsd1}.
We can see that the scaling formula
(\ref{eq:pofdscaling}) is satisfied from the time 8 to
6144.
The data for longer times are not shown here because after 10000
independent runs they still have
too big noise.
The scaling function $F(x)$ has a maximum again
near $x=0.4$. 
This indicates creation of a quasi-regular domain
structure, but the distribution has very broad tail for larger $x$
extending up to $x\simeq 4$, which makes the domain structure
irregular. 
We compared the scaling for different $K$. For times 
shorter than the saturation time the scaling function does not depend
on $K$, which is demonstrated in the inset in Fig. \ref{fig:dsd1}
for $K=1.0, 2.0$, and $3.0$.

The behavior similar to that in the strong coupling regime
was observed for the kinetic Ising
model in one dimension. Numerical simulations of the Ising model
at zero temperature 
\cite{..ising..} show that the distribution of domain sizes  obeys the form
(\ref{eq:pofdscaling}) with a peak around $d/D\simeq
0.4$. Also the form of the function $F(x)$ found in \cite{..ising..} looks
very similar to our results. On the contrary, for any non-zero
temperature it may be shown analytically, that the distribution of
domain sizes in equilibrium,  i.e.  in the infinite time limit,
is exponential (see  e.g.  \cite{baxter82}).
Analytical results for the zero-temperature kinetic
Ising model \cite{de_zei_96} give the asymptotic results 
$F(x) \simeq \pi x$ for $x\to 0$
and $F(x) \simeq \exp(-Ax+B)$ for $x\to\infty$, with
$A=\frac{1}{2}\zeta(3/2) = 1.30618...$, and $B=0.597...$, which is in
good agreement with our results, as Fig. \ref{fig:dsd1} shows.

Summing up, we should again
distinguish two time regimes.  At initial times, the form of the
distribution of
domain sizes as a function of $x$ does not change and it is
characterized by a pronounced peak. 
Scaling (2) holds and the only change during the time
evolution is the increase of the average domain size $D$. 
When $D$ begins to saturate, the peak vanishes and in
the saturated regime, the distribution of domain sizes is
exponential. 
The reason why in the case
of strong coupling
($K=2.0, 3.0$) the peak remained for all times observed is simply that the
duration of the simulation is still much shorter than the saturation time.

\subsubsection{Correlation function}
The correlation function
$S(r,t)=\frac 1L\sum_{i=1}^L
\langle \sigma\left( i+r,t\right)
\sigma\left( i,t\right)\rangle$
decays nearly exponentially with the distance $r$.
The decay is characterized by the correlation length
$\xi^s(t,K)$ which we computed 
by fitting the data to the function
$\exp(r/\xi^s(t))$ in an interval $(0,r_s)$ in which the decay is
actually well described by an exponential. 
According to our experience a good recipe for fixing 
the interval $(0,r_s)$ is to find the
distance $r_s(t)$ as the minimal distance
for which $S(r_s(t),t) \le 0.05$.
This is done for given time and coupling.

The time behavior of the correlation length $\xi^s(t,K)$ is 
similar to that of the average domain size $D(t,K)$.
It increases with time as a power law
$\xi^s(t,K)\propto t^{\kappa_{\rm eff}(K)}$
(we found $\kappa_{\rm eff} \simeq
0.5$ for $K=2$). For weak coupling $K$, 
$\xi^s(t,K)$ saturates to a finite value $\xi^s_{\rm sat}$ 
at about the same time $t^{(D)}_{\rm
sat}$ as the saturation 
of the average surface domain size occurs. The saturation for large $K$'s
cannot be seen because the simulation time is insufficient.
The dependence of the saturated correlation
length $\xi^s_{\rm sat}$ on $K$ is shown in Fig. \ref{fig:dep-on-K}. 
It can be fitted by an 
exponential, $\xi^s_{\rm sat} \simeq {\rm e}^{(3.29 \pm 0.03)K}$. 
This behavior is similar to what was observed for
for $D_{\rm sat}$ (cf. Fig. \ref{fig:dep-on-K}).

\begin{figure}[hb]
\centering
\vspace*{90mm}
\includegraphics{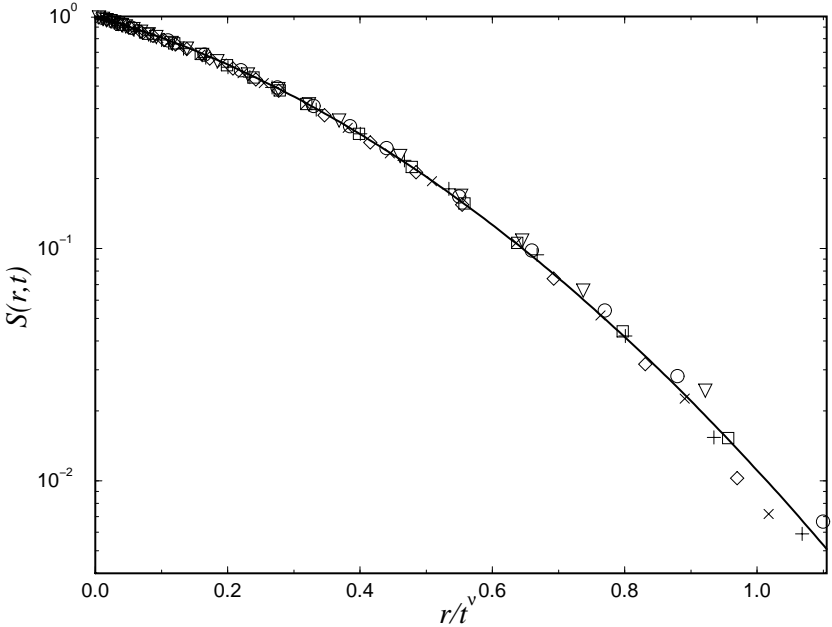}
\caption{
Spin-spin correlation function $S(r,t)$ as function of
$\frac{r}{t^{\nu}}$
 for $K=2$ in times 
512 ($\circ$),
1024, ($\Box$),
6144 ($\Diamond$),
32768 ($\times$),
65536  ($\triangle$),
131072  ($+$).
The solid line is the fit
with function $\exp(-bx-ax^2)$, $a=2.6$, $b=1.86$, $x=r/t^{\nu}$, $\nu=0.465$.
}
\label{fig:spin-spin}
\end{figure}

For times shorter than $t^{(D)}_{\rm sat}$ we observed the following
scaling: $S(r,t)=\bar{S}(r/t^\nu)$,
the value of exponent $\nu$  for $K=2$ is $\nu = 0.465$
 (Fig. \ref{fig:spin-spin}).
The
function $\bar{S}(x)$ was found to have the form $\bar{S}(x) =
\exp(-bx-ax^2)$ where the parameters $a,b$ were fitted as
$a=2.6\pm0.1$ and $b=1.86\pm0.09$.
This behavior agrees with analytical results 
for the zero-temperature kinetic Ising model \cite{bray89}. 
The correlation functions $S(r,t)$ for time larger than $t^{(D)}_{\rm sat}$
do not depend on time but they depend on $K$.
These functions $S_{\rm sat}(r,K)$
can be also scaled into universal form using
the saturated correlation length 
$S_{\rm sat}(r,K)=\tilde{S}(r/\xi^s_{\rm sat}(K))$.

\subsubsection{Concentration of components}
\label{sec:magnetization}
We measured also the quantity $m=\frac{1}{L}\sum_i \sigma_i$,
which is an analog of the surface magnetization
and from which the concentrations of both components,
$c_{+}=\frac{1+m}{2}$, $c_{-}=\frac{1-m}{2}$, 
on the surface can be determined.
In all the simulations described so far the external field was zero,
therefore  $m$ (after a sufficient averaging) was zero, i.e.
average concentration of one component was {50 \%} and constant in time.
However, for nonzero field  average $m$
is no longer zero and it depends on time and external field,  i.e.
 there is time dependent excess of one component.
When time of the simulation is sufficiently long
we arrive to the stationary state with a characteristic value of
$m^{\rm stat}(H,K)$ which corresponds to the concentration of
both components produced in the stationary regime. 

The quantity $m$ can be obtained from our simulations, 
however, it is the subject of rather strong fluctuations.
Instead of $m$ 
we measured the corresponding bulk quantity $M$.
In order to follow the time dependence we measured $M$
in the same time intervals as the surface width, i.e. in the logarithmic scale,
but we recorded only sum of $\sigma$'s of particles deposited 
between two intervals of measurement divided by the total number
of particles deposited during this interval.
Therefore, $M(t)$ is not the total bulk magnetization
but ``incremental'' bulk magnetization.
The numbers of particles deposited in each time interval
increase as $L$, $2L$, $4L$, $8L$, $\cdots$.  
Hence, for a long time we are averaging over many bulk particles
and approaching to the total bulk magnetization.
The fluctuations of $M(t)$ are smaller than fluctuations of $m(t)$.
In all simulations in this subsection we started from the substrate
with alternating types of particles, it implies $M(0)=m(0)=0$.
We continued the simulations up to time when the stationary regime was reached
and we measured $M^{(\rm stat)}(H,K)$.
Time needed depends on both $H$ and $K$. 
From $M^{(\rm stat)}$ we can 
obtain the stationary concentrations.
However, we prefer to use the quantity $M^{(\rm stat)}$ in order
to utilize the analogy with magnetic systems. 
\begin{figure}[hb]
\centering
\vspace*{90mm}
\includegraphics{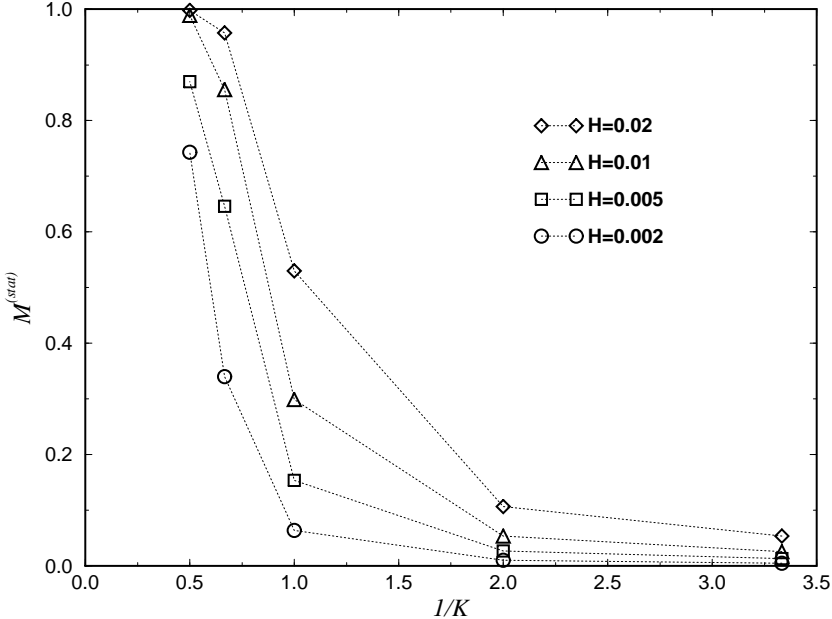}
\caption{Dependence of stationary bulk magnetization 
$M^{\rm stat}$ on inverse coupling $\frac{1}{K}$
for several external fields.
}
\label{fig:mag}
\end{figure}
In Fig. \ref{fig:mag}
 we plotted   $M^{(\rm stat)}(H,K)$ 
as function of $1/K$ for several values of $H$. 
When coupling is small the effect of external field is weak
and $M^{(\rm stat)}$ is close to zero, 
 i.e. concentration changes very little,
and $M^{(\rm stat)}$ is close to 
the value corresponding to zero coupling \cite{zeroK}.
For large $K$, dependence on external field
becomes very strong (cf. also Fig.  \ref{fig:conf}).

For $K> 1.5$ the magnetization seems to go to a finite limit when
external field goes to zero indicating possible existence 
of the bulk phase transition.
However, our present data are for rather small system and 
more detailed study of 
finite-size effects is needed in order to decide
on the presence or absence of the
bulk phase transition.

\section{Discussion}
\label{sec:dis}
The crossover in the surface width as well as in the height-difference
correlation function is clearly related to stopping of phase ordering
on the surface.
We observed that time for saturation of domain $t^{(D)}_{\rm sat}$
is approximately proportional to time $t_{\rm cross}$
for crossover to the KPZ exponent $\beta=1/3$
in evolution of the surface width
(Fig. \ref{fig:dep-on-K}).
Due to progressively increasing time and system size needed for the simulation
we cannot decide from our data whether the crossover is present for
any coupling, or if there is a phase transition at a certain critical
coupling $K_c$, and for $K>K_c$ the exponents remain enhanced.
However, we do believe that the crossover is present for
any value of $K$, but it is hard to see it for strong coupling
because $t_{\rm cross}$ is larger than possible simulation time.

In the surface morphology, the most striking features are the
pyramids or teeth observed for a sufficiently large coupling $(K\ge 1.5)$. 
This can be understood from the rules of growth. 
Let us consider a growth site with all nearest neighbors
of the same type.
The probability that it will be occupied by a new particle is 
$\propto (e^{3K}+e^{-3K})$. 
The first term, $e^{3K}$ gives the probability
that a new particle will be of the same type as the old ones. 
This is much
larger than the second term corresponding to the probability
that a new particle will be
 of the opposite type. The creation of new 
domain walls is thus strongly inhibited and growth proceeds preferably
by adding the particles of the same type.
On the other hand,
probability to occupy a growth site next to the boundary between two domains is
$\propto (e^{K}+e^{-K})$, i.e. smaller than the growth probability inside
the domain. 
Hence, growth inside the domain is preferable.
This leads
to the formation of pyramid-like features
composed of one type of particles with facets of maximal slope
and domain walls in bottoms of the valleys.
In other words growth is pinned by domain walls. This is in accord
with the results on non-homogeneous growth, where the presence of the
inhomogeneity leads to the formation of a dip in the surface 
\cite{wo_ta_90,sla_ko_97}.

At the same time, if we observe the deposition of particles next to
the domain wall, we can see that the wall movement is due to
deposition of opposite type of particle, which has probability $\propto
{\rm e}^{-K}$, while no movement has probability $\propto {\rm
e}^K$. That is why the wall movement is very slow for large $K$.
The surface domains with slowly moving walls result in long vertical
lamellae (cf. Fig. \ref{fig:conf}).
This leads to at first sight surprising fact that the width of lamellae
of different types of particles decreases with coupling (for equal
times),
which we observed for the growth with initial condition fixed by
neutral substrate
\cite{A-vs-B}.
This is also reason of longer transient times for larger $K$.

Non zero external field leads to surface (as well as bulk)
magnetization, or in the context of alloy growth, to changing stoichiometry.
We can still define exponent $\psi$ for growth of the dominant domain
size as well as exponents for kinetic roughening.
Effect of this symmetry breaking on values of exponents remains to be
studied.

\section{Conclusion}
\label{sec:conc}
We have investigated the interplay between phase ordering and kinetic 
roughening using the 1+1 dimensional two-component
single-step SOS growth model.
We examined validity of scaling for both phenomena
and measured the effective scaling exponents.

We observed two 
situations depending on the strength of coupling $K$
between two types of particles.
For a moderate, sufficiently strong ($K>0.3$) but not very large
($K<1.8$) coupling, there is crossover in time and spatial behavior
of geometrical characteristics of the surface profile.
The effective exponents $\alpha$ and $\beta$ 
for time shoter than $t_{\rm cross}$ are
significantly larger than the KPZ exponents.
After crossover we observed 
the KPZ exponents.
Surface ordering
proceeds only up to a finite time $t^{D}_{\rm sat}(K)$ after which it stops.
Crossover time $t_{\rm cross}$ is proportional to $t^{D}_{\rm sat}(K)$
and both are exponentially increasing with the strength of coupling.
For strong coupling ($K\ge 2$), we observed only enhanced exponents
and ordering continued during all time of our simulation.
However, we believe that this difference is only due to
finite time of our simulation, and that 
ordering will eventually stop and the crossover to the KPZ behavior will 
occur for any coupling.

The intermediate growth regime is
connected with surface phase ordering.
It results in enhanced and more rapid kinetic roughening.
There is also crossover
in geometrical characteristics with increasing coupling
for fixed time and length scale.
Scaling exponents are continuously increasing with $K$.
We found
that, for sufficiently strong coupling, scaling with enhanced
exponents is satisfied over many decades.

During phase ordering
the average size of the surface domains $D$ increases in time as 
$D\propto t^{\psi}$  with the exponent $\psi$ close to $1/2$.
The spin-spin correlation function and the distribution
of domains obey scaling with the same exponent.
Our results for the surface ordering in the intermediate regime are in
agreement with 
the known results for one dimensional  kinetic Ising model 
with nonconserved order parameter at zero temperature.
We expect that 
the phase ordering on the surface is essentially described by the kinetic
Ising model for any coupling.
Domain growth stops when the average domain size
reaches the equilibrium correlation length.
This is reflected in turn by crossover in effective exponents for kinetic
roughening.

Our results lead to strong belief that in 1+1 dimensions there is no
new universal behavior and that the TCSS model belongs 
to the KPZ universality class for any value of coupling.
However, since the crossover time and the correlation length
are increasing exponentially with coupling,
the new behavior in the intermediate regime
can be dominant for practically relevant times
and length scales.

It is of interest to study the TCSS model in 2+1 dimensions.
If the analogy with the kinetic Ising model is valid also here,
the size of surface domains will not be  restricted
and for some critical value of $K$, it will diverge.
Then a new universal behavior may be observed.
Furthermore, one can expect that in 2+1 dimensions 
a phase transition in kinetic roughening exists.
It would be also desirable to explore 
scaling in different growth models
for binary systems, in particular in models with the surface diffusion.

\begin{center}
{\bf 
Acknowledgments}\\
\end{center}
This work was
supported by grants No. A 1010513 of the GA AV \v{C}R and  
No. 202/96/1736 of the GA \v{C}R.


\begin{references}
\bibitem{flat}To be more precise the initial configuration
is $h^{(o)}_i=a \cdot (i\; {\rm mod}\; 2)$ 
because of the single-step constraint.
Here $a$ is the lattice constant.
The initial configuration is not strictly flat
but it has the intrinsic width $w^{\rm int}=\frac{a}{4}$.
\bibitem{additional}
The additional structure seen for initial times in the case of
large couplings is the transient caused by the initial conditions.
When we start from a different initial composition of the substrate
the dependence of $w$ on $t$ is different for some initial time interval.
This interval is increasing with coupling, for $K=3$ it is about
100 ML (cf. Fig. \ref{fig:rough}).
\bibitem{kineticIM}
The kinetic Ising model evolves towards a stationary state which is
an equilibrium state of the thermodynamical Ising model.
The domain size $D$ in the Ising model increases as $D\propto e^{2K}$.
\bibitem{zeroK}In the case of zero coupling, particles of 
type $+$ and $-$ are deposited with  probabilities 
$p_{+}=\frac{e^{H}}{e^{H}+e^{-H}}$ and $p_{-}=\frac{e^{-H}}{e^{H}+e^{-H}}$  
respectively. It results in magnetization $m=p_{+}-p_{-}=\tanh H.$
\bibitem{A-vs-B}
In this respect we would expect very
different behavior of the variant B of the model (Fig. \ref{fig:sssos}b), 
where the
ambiguity of the deposited particle type favors the domain wall
movement. Indeed, in the simplified case when the deposition next to
the domain wall is equally probable as elsewhere, the correlation time
grows as $t_c\propto {\rm e}^{2K}$ for the variant A, while it is
independent of $K$ for the variant B.
\end{references}

\end{document}